\documentclass[aps,prl,reprint,superscriptaddress,twocolumn]{revtex4-2}

\usepackage{amsmath,amssymb,amsfonts}
\usepackage{color}
\usepackage{bm}
\usepackage{graphicx}
\usepackage{braket}
\usepackage{lineno}
\usepackage{ulem}

\begin{document}

\title{Interaction-driven spin polaron in itinerant flat-band ferromagnetism}

\author{Wei-Tao \surname{Zhou}}
\affiliation{National Laboratory of Solid State Microstructures and Department of Physics, Nanjing University, 210093 Nanjing, China}

\author{Zhao-Yang \surname{Dong}}
\email{zhydong@njust.edu.cn}
\affiliation{School of Physics, Nanjing University of Science and Technology, Nanjing 210094, China}
\affiliation{Jiangsu Physical Science Research Center, Nanjing 210093, China}

\author{Jian-Xin \surname{Li}}
\email{jxli@nju.edu.cn}
\affiliation{National Laboratory of Solid State Microstructures and Department of Physics, Nanjing University, 210093 Nanjing, China}
\affiliation{Collaborative Innovation Center of Advanced Microstructures, Nanjing University, 210093 Nanjing, China}
\affiliation{Jiangsu Key Laboratory of Quantum Information Science and Technology,
Nanjing University, Suzhou 215163, China}
\date{\today}

\begin{abstract}
Interaction effects are dramatically enhanced in flat-band systems due to quenched kinetics, facilitating the binding of single excitations into composite quasiparticles.
In this work, we present a comprehensive study of spin polarons over the entire momentum and energy space within the Mielke-Tasaki model using projected exact diagonalization.
We identify distinct low-energy spin polarons at momenta $q=0$ and $q=\pi$, and also find multiple high-energy branches of spin polaron.
It is demonstrated that the interaction-induced Hartree dispersion plays a decisive role in determining the momentum sector of low-energy spin polarons.
Furthermore, by introducing a finite bandwidth, we unravel the underlying binding mechanisms: the formation of low-energy spin polarons is governed by the conventional virtual exchange mechanism, whereas the high-energy spin polarons arise from a joint effect of the effective attraction and virtual exchange.
Our results suggest promising avenues for realizing spin polaron crystals and exploring novel superconducting pairing mechanisms in moir\'e materials like twisted $\text{WSe}_{2}$ and $\text{MoTe}_{2}$.
\end{abstract}
\maketitle
\textit{Introduction.---}
Itinerant flat-band ferromagnetism (IFBFM) has attracted significant interest in condensed matter physics driven by the recent surge of research in moir\'e materials\cite{Cao2018a,Cao2018,Oh2021,Xie2021,Park2023,Xu2023,Anderson2023,Cai2023,Zeng2023,Wang2024,Lu2024,Wang2024,Bernevig2025,Park2025,Han2025}.
In contrast to spin systems, the unfrozen charge degree of freedom in itinerant electron systems facilitates rich spin-charge coupling, giving rise to novel physical phenomena.
Due to the quenched kinetics, the interactions in flat-band systems are greatly enhanced. This enhancement tends to bind single excitations, such as electron ($e$), hole ($h$) and spin ($\sigma^{-}$) excitations to form composite quasiparticle, such as excitons, trions and spin polarons, emerging from the $eh$, $eeh$/$ehh$ and $e\sigma^{-}$ excitations, respectively.

The well-established observation of excitons in moir\'e materials\cite{Yu2017,Tran2019,Jin2019,Alexeev2019,Seyler2019,Shimazaki2020,Forg2021,Schmitt2022,Naik2022,Karni2022,Xiong2023,Zhao2023}, together with the more recent detection of trions in twisted $\text{MoSe}_2/\text{WSe}_2$\cite{Nguyen2025,Qi2025} and $\text{MoTe}_2$\cite{Anderson2024} have further established flat-band systems as promising platforms for exploring novel quasiparticles as well as hosting exotic correlated phases of matter.
Beyond the extensively studied charge degrees of freedom, the rich landscape of spin-charge coupled quasiparticles, such as spin polarons, remains largely uncharted.
Investigating spin-charge coupled quasiparticles is pivotal for understanding magnetic quantum phase transitions, exploring exotic superconducting pairing mechanisms, and advancing electromagnetic manipulation in doped strongly correlated systems.

In this work, we study spin polarons in a prototypical flat-band model---Mielke-Tasaki Model, which can stabilize IFBFM at half filling of the flat band under sufficiently weak interaction. 
The adoption of this one-dimensional model can allow us to reliably unveil the underlying physical mechanisms through large-scale numerical calculations. The findings reported here hold generality within flat-band ferromagnetic systems.
The emergence of the {\it local} spin polarons has been predicted in Ref.\cite{Prembabu2025} using finite DMRG, which identify the formation of quasiparticle by comparing the energy of the composite excitation to that of its free constituents.
Specifically, the energies of $e$, $\sigma^{-}$ and $e\sigma^{-}$ excitations is $E_{e} = E(N,1)-E(N,0)$, $E_{\sigma^{-}} = E(N-1,1)-E(N,0)$ and $E_{e\sigma^{-}} = E(N-1,2)-E(N,0)$ [$E(N_{\uparrow}, N_{\downarrow})$ denotes the ground state energy in subspace $(N_{\uparrow}, N_{\downarrow})$], respectively, and the formation of the local spin polaron is identified by $E_{e\sigma^{-}} < E_{e} + E_{\sigma^{-}}$.
However, this criterion entails two limitations: (a) it neglects momentum conservation during the binding process, thereby failing to capture the momentum-dependent formation of spin polarons across the entire BZ; (b) it cannot identify the spin polarons embedded within the continuum or situated in the high-energy region.

Here, we employ a combination of single mode approximation and projected exact diagonalization (PED) to identify the spin polaron over the entire momentum and energy space.
We present the spectrum of the $e_{\downarrow}e_{\downarrow}h_{\uparrow}$ excitation over the entire BZ and identify the spin polarons out of the continua both by the binding energy and spectral.
We find two types of spin polarons locating around momenta $q=0$ and $q=\pi$, whose emergence is related to the Hartree dispersion of the bare spin-$\downarrow$ electron.
We additionally unveil multiple branches of spin polarons in the high-energy region, which indicates the binding of the bare electron with optical magnons.
Finally, we elaborate that the formation mechanism of low-energy spin polarons is attributed to the virtual exchange while that of high-energy spin polarons to a joint effect of the effective attraction and virtual exchange.
Our work is expected to deepen the understanding of flat-band quasiparticles and motivate the related experimental search, while also paving the way for electromagnetic engineering via spin-charge control.
\begin{figure}[htbp]
    \centering
    \includegraphics[width=\linewidth]{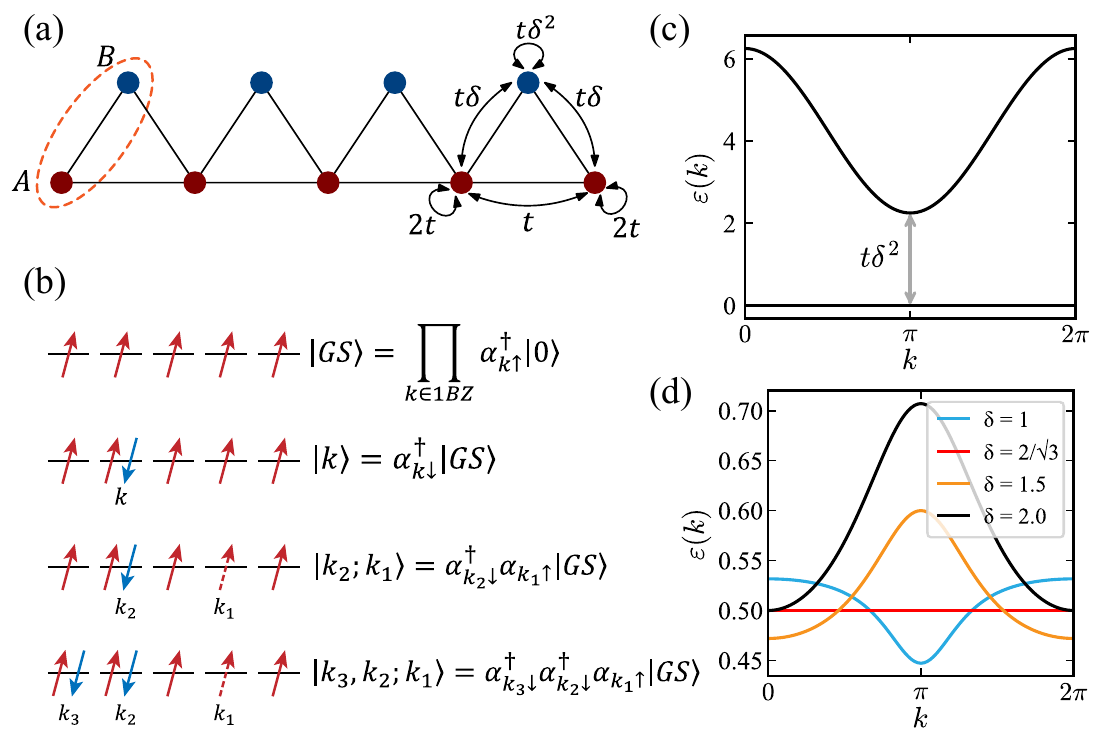}
    \caption{
    Model and single-particle dispersion.
    (a) Mielke-Tasaki Model.
    (b) Sketches of the flat-band ferromagnetic ground state, the $e_{\downarrow}$, $e_{\downarrow}h_{\uparrow}$ and $e_{\downarrow}e_{\downarrow}h_{\uparrow}$ excitations.
    (c) Band structure of Mielke-Tasaki model.
    (d) Hartree band of a spin-$\downarrow$ electron in the flat-band ferromagnetic state.
    }
    \label{fig1}
\end{figure}

\textit{Model and Method.---}
The Mielke-Tasaki model exhibits an exact flat band by fine tuning the hopping parameters to satisfy the destructive interference criterion, thereby stabilizing compact localized states\cite{Mielke1991,Mielke1991a,Mielke1992,Tasaki1992,Mielke1993,Tasaki1994,Tasaki1998,Tasaki1998a,Mielke1999,Maimaiti2017}, whose Hamiltonian is [sketched in Fig.\ref{fig1}(a)]:
\begin{equation}
    H = t\sum_{i\sigma} \psi^{\dagger}_{i\sigma} \psi_{i\sigma} + \sum_{in} U_{n} n^{\uparrow}_{in} n^{\downarrow}_{in}
    \label{eq1}
\end{equation}
where $\psi_{i\sigma}=c_{iA\sigma} + \delta c_{iB\sigma} + c_{i+1,A\sigma}$, $\sigma$ and $n$ label the spin and sublattice, respectively, and $U_{n}$ is the Hubbard interaction on the $n$ sublattice.
We project Eq.(\ref{eq1}) into its flat band, which yields:
\begin{equation}
    \mathcal{P} H \mathcal{P} = \sum_{nq} U_{n} \rho^{\downarrow}_{qn} \rho^{\uparrow}_{-qn}
    \label{eq2}
\end{equation}
where $\rho^{\sigma}_{qn} = 1/\sqrt{N}\sum_{k} M^{\sigma}_{n}(k,q)\alpha^{\dagger}_{k+q\downarrow}\alpha_{k\uparrow}$,  $\alpha^{\dagger}_{k\sigma}(\alpha_{k\sigma})$ creates(annihilates) a spin-$\sigma$ electron with momentum $k$ on the flat band, $M^{\sigma}_{n}(k,q)=u^{\sigma*}_{k+q,n} u^{\sigma}_{kn}$ and $u^{\sigma}_{kn}$ is the unitary transformation coefficience  ($\alpha^{\dagger}_{k\sigma}=u^{\sigma}_{kA}c^{\dagger}_{kA\sigma}+u^{\sigma}_{kB}c^{\dagger}_{kB\sigma}$).

We first consider the $e_{\downarrow}$ and $e_{\downarrow}h_{\uparrow}$ excitations, which are the constituent excitations of the spin polaron out of the $e_{\downarrow}e_{\downarrow}h_{\uparrow}$ excitation, sketched in Fig.\ref{fig1}(b).
In single mode approximation (although no approximation after projection), the $e_{\downarrow}$ and $e_{\downarrow}h_{\uparrow}$ excitation on the flat-band ferromagnetic state is encoded in subspace $\ket{k} \equiv \alpha^{\dagger}_{k\downarrow}\ket{\text{GS}}$ and $\ket{k_{2};k_{1}} \equiv \alpha^{\dagger}_{k_{2}\downarrow} \alpha_{k_{1}\uparrow} \ket{\text{GS}}$, respectively, where $\ket{\text{GS}} \equiv \prod_{k} \alpha^{\dagger}_{k\uparrow} \ket{0}$ is the spin-$\uparrow$ flat-band ferromagnetic ground state.
The spectrum of the $e_{\downarrow}$ excitation can be obtained by diagonalizing Eq.(\ref{eq2}) in subspace $\ket{k}$, which yields:

\begin{equation}
    \varepsilon^{\downarrow}_{e}(k) = U_{A}n^{\uparrow}_{A}n^{\downarrow}_{kA} + U_{B}n^{\uparrow}_{B}n^{\downarrow}_{kB}
    \label{eq3}
\end{equation}
where $n^{\sigma}_{kn}=M^{\sigma}_{n}(k,0)$ is the occupation of a flat-band electron with spin $\sigma$ and momentum $k$ at sublattice $n$, and $n^{\uparrow}_{A/B}=1/N \sum_{k} n^{\uparrow}_{k,A/B}$. 
$\varepsilon^{\downarrow}_{e}(k)$ can be interpreted as the energy of the bare spin-$\downarrow$ electron in the effective field of the ferromagnetic ground state, thus dubbed the interaction-induced Hartree energy.
It can be fine tuned to vanish by adjusting $\delta$ or $U_{B}/U_{A}$ to satisfy the condition $U_{A}n^{\uparrow}_{A}=U_{B}n^{\uparrow}_{B}$, which is $U_{B}/U_{A}=\delta/(\sqrt{\delta^{2}+4}-\delta)$.
When $U_{B}/U_{A}=1$, the Hartree dispersion vanishes at $\delta_{c}=2/\sqrt{3}$, as shown in Fig.\ref{fig1}(d).
The Hartree dispersion is essential for the emergence of the spin polaron, which will be shown in the following discussion.

The $e_{\downarrow}h_{\uparrow}$ excitation is determined by the following equation:
\begin{equation}
\begin{split}
    \mathcal{P} H \mathcal{P} \ket{k_{2};k_{1}} &= \varepsilon^{\downarrow}_{e}(k_{2})\ket{k_{2};k_{1}} \\ &- \sum_{q} \mathcal{S}(k_{1},k_{2},q) \ket{k_{2}+q;k_{1}+q}
\label{eq4}
\end{split}
\end{equation}
where the two terms are the Hartree and exchange term, respectively.
$\mathcal{S}(k_{1},k_{2},q) = 1/N \sum_{n} U_{n} M^{\uparrow}_{n}(k_{1}+q,-q)M^{\downarrow}_{n}(k_{2},q)$ is the strength of scattering from $\ket{k_{2};k_{1}}$ to $\ket{k_{2}+q;k_{1}+q}$.
Due to the translational symmetry, this subspace can be block diagonalized by total momentum $Q = k_{2} - k_{1}$.
Similarly, the $e_{\downarrow}e_{\downarrow}h_{\uparrow}$ excitation can be derived by:

\begin{equation}
\begin{split}
    \mathcal{P} H \mathcal{P} \ket{k_{3},k_{2};k_{1}} &= (\varepsilon^{\downarrow}_{e}(k_{2}) + \varepsilon^{\downarrow}_{e}(k_{3})) \ket{k_{3},k_{2};k_{1}} \\ & - \sum_{q} \mathcal{S}(k_{1},k_{2},q) \ket{k_{3},k_{2}+q;k_{1}+q} \\ & - \sum_{q} \mathcal{S}(k_{1},k_{3},q) \ket{k_{3}+q,k_{2};k_{1}+q}
\end{split}
\label{eq5}
\end{equation}
where $\ket{k_{3},k_{2};k_{1}} \equiv \alpha^{\dagger}_{k_{3}\downarrow} \alpha^{\dagger}_{k_{2}\downarrow} \alpha_{k_{1}\uparrow} \ket{\text{GS}}$, the first term is the Hartree energies of the two spin-$\downarrow$ electrons and the last two terms are the scattering from $\ket{k_{3},k_{2};k_{1}}$ to $\ket{k_{3},k_{2}+q;k_{1}+q}$ and $\ket{k_{3}+q,k_{2};k_{1}+q}$, respectively, and this subspace can be block diagonalized by total momentum $Q = k_{2} + k_{3} - k_{1}$.

\begin{figure}[htbp]
    \centering
    \includegraphics[width=\linewidth]{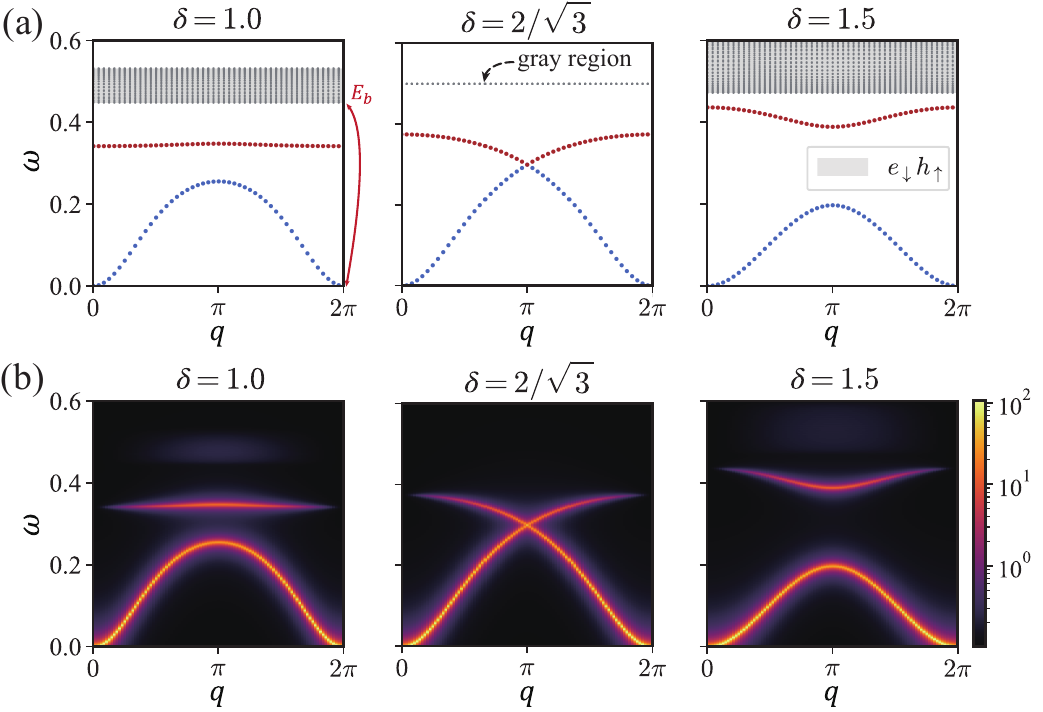}
    \caption{Spectrum and spectral function $A(q,\omega)$ of the $e_{\downarrow}h_{\uparrow}$ excitation.
    (a) Spectrum of the $e_{\downarrow}h_{\uparrow}$ excitation at different $\delta$.
    The dots are the results obtained from PED and the shaded gray region is the $e_{\downarrow}h_{\uparrow}$ continuum obtained by simply summing the energies of the bare constituents.
    The blue and red dispersion is the acoustic and optical magnon, respectively.
    The red arrow denotes the binding energy of the acoustic magnon.
    (b) Spectral function of the $e_{\downarrow}h_{\uparrow}$ excitation with quasiparticle operator $\beta^{\dagger}_{i} = \mathcal{P}\sum_{n}c^{\dagger}_{in\downarrow}c_{in\uparrow}\mathcal{P}$.
    All panels are calculated at $U_{A}=U_{B}=1$.
    }
    \label{fig2}
\end{figure}

\textit{Results.---}
We first fix $U_{A}=U_{B}=1$ and inspect these excitations by varying $\delta$.
In order to character and distinguish the quasiparticles from the continuum, we also calculate the spectral function of quasiparticles $A(q,\omega)=-(1/\pi) \text{Im}G(q,\omega)$ with:
\begin{equation}
    G(q,\omega) = \sum_{m} \frac{|\bra{\psi_{mq}} \beta^{\dagger}(q) \ket{\text{GS}}|^{2}}{\omega - \omega_{mq} + i\delta}
    \label{eq6}
\end{equation}
where $\omega_{mq}$, $\psi_{mq}$ is the eigen energy and eigen state at the total momentum $q$, which is obtained by PED, and $\beta^{\dagger}(q)$ is the quasiparticle operator corresponding to the specified excitation discussed.

The spectrum and spectral function of the $e_{\downarrow}h_{\uparrow}$ excitation are shown in Fig.\ref{fig2}.
In contrast to conventional spin systems, the itinerant magnon ($\sigma^{-}$) in flat-band electronic systems emerges as a quasiparticle out of the $e_{\downarrow}h_{\uparrow}$ excitation, bound by an effective attraction between $e_{\downarrow}$ and $h_{\uparrow}$\cite{Su2018,Su2019,Gu2021,Zhou2025,Qiu2025}.
In Fig.\ref{fig2}(a), the $e_{\downarrow}h_{\uparrow}$ continuum
is clearly recognized as the shaded gray region, plotted by a summation of the energies of the individual $e_{\downarrow}$ and $h_{\uparrow}$ excitations (the latter is zero), which coincides with the high-energy spectrum obtained by PED. Below the $e_{\downarrow}h_{\uparrow}$ continuum, two distinct branches of magnons—acoustic ($\sigma^{-}_{a}$) and optical ($\sigma^{-}_{o}$) branches emerge, which are represented by the blue and red lines, respectively. As $\delta$ is varied, the width of the $e_{\downarrow}h_{\uparrow}$ continuum first shrinks, vanishes at $\delta=\delta_{c}$, and broadens again. This behavior tracks the evolution of the Hartree dispersion $\varepsilon^{\downarrow}_{e}(k)$, as illustrated in Fig.\ref{fig1}(d).
In this process, two magnons undergo a band inversion and touch at $q=\pi$, suggesting a topological transition as has been discussed before\cite{Su2018}.
To reflect the bound-state nature of the magnon, we study the spectral function corresponding to the quasiparticle operator $\beta^{\dagger}_{i} = \mathcal{P}\sum_{n}c^{\dagger}_{in\downarrow}c_{in\uparrow}\mathcal{P}$, and the results are shown in Fig.\ref{fig2}(b). It reveals that the spectral weight is predominantly concentrated on the two magnons, with negligible weight on the $e_{\downarrow}h_{\uparrow}$ continuum.

\begin{figure}[htbp]
    \centering
    \includegraphics[width=\linewidth]{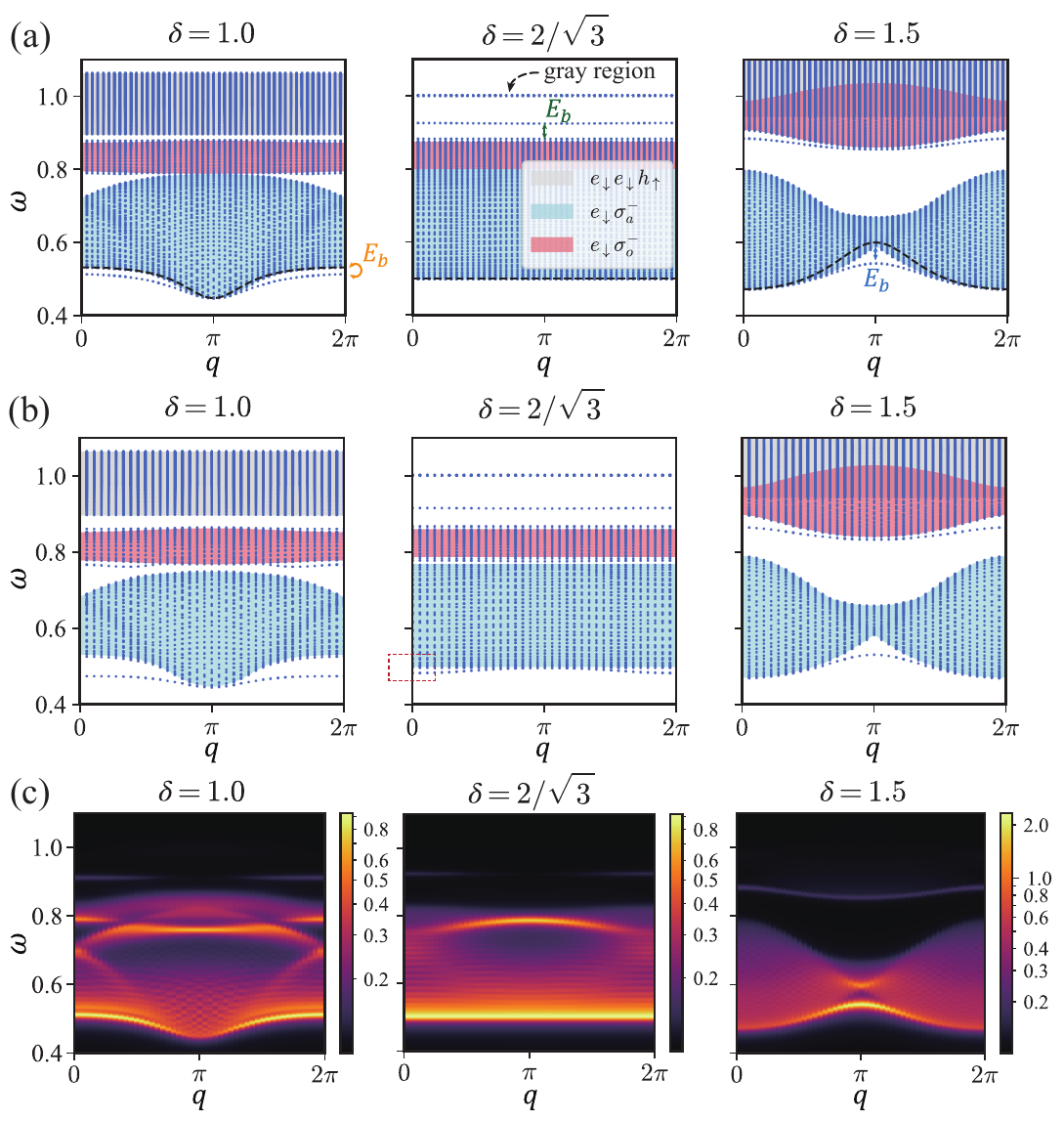}
    \caption{Spectrum and spectral function $A(q,\omega)$ of the $e_{\downarrow}e_{\downarrow}h_{\uparrow}$ excitation.
    (a) Spectrum of the $e_{\downarrow}e_{\downarrow}h_{\uparrow}$ excitation at different $\delta$.
    The dots are the results obtained from PED and the shaded gray, red and blue regions are the $e_{\downarrow}e_{\downarrow}h_{\uparrow}$, $e_{\downarrow}\sigma^{-}_{o}$ and $e_{\downarrow}\sigma^{-}_{a}$ continua, respectively.
    The black dashed lines are the bare spin-$\downarrow$ electron Hartree dispersions.
    Three colored arrows denote the binding energies of the low-energy $q=0$, $q=\pi$ and high-energy spin polaron, respectively.
    (b) Spectrum of the $e_{\downarrow}e_{\downarrow}h_{\uparrow}$ excitation after including the inter-band mixing in the PED. The red dashed block shows the emergence of the spin polaron at $\delta=\delta_{c}$.
    (c) Spectral function of the $e_{\downarrow}e_{\downarrow}h_{\uparrow}$ excitation with quasiparticle operator $\beta^{\dagger}_{i}=\mathcal{P}c^{\dagger}_{iA\downarrow}c_{iA\uparrow}c^{\dagger}_{iB\downarrow}\mathcal{P}$.
    All panels are calculated at $U_{A}=U_{B}=1$.
    }
    \label{fig3}
\end{figure}

Having analyzed the bare constituents, we now investigate the spin polaron out of the $e_{\downarrow}e_{\downarrow}h_{\uparrow}$ excitation.
As shown in Fig.\ref{fig3}(a), the shaded blue, red, and gray regions correspond to the $e_{\downarrow}\sigma^{-}_{a}$, $e_{\downarrow}\sigma^{-}_{o}$, and $e_{\downarrow}e_{\downarrow}h_{\uparrow}$ continua, respectively.
These regions are determined by summing the energies of the corresponding bare constituents and also consist with the spectrum obtained by PED.
Below the $e_{\downarrow}\sigma^{-}_{a}$ continuum, distinguishable spin polarons emerge around $q=0$ for $\delta=1.0$ and $q=\pi$ for $\delta=1.5$.
The black dashed lines are the bare spin-$\downarrow$ electron Hartree bands.
Comparing the dispersion of spin polaron with Hartree electron, we find that these low-energy spin polarons mimic the Hartree dispersions of bare electrons, albeit shifted to a slightly lower energy.
It indicates that the low-energy spin polaron essentially consists of an electron dressed by an acoustic magnon with a renormalized energy, thereby manifesting strong spin-charge coupling.
This characteristic enables the cross-control of electrical and magnetic degrees of freedom, consequently promoting the development of electromagnetic engineering.
At the critical point $\delta_{c}=2/\sqrt{3}$, the spin polaron merges into the continuum, making it energetically indistinguishable.
We attribute this phenomenon to the vanishing of the electron's Hartree dispersion.
Under this circumstance, the quenched Hartree kinetics is insufficient to sustain the virtual exchange processes required for the electron to form a stable bound state with the magnon.
If the inter-band mixing is taken into account, the dispersive upper band can provide additional kinetics, allowing the electron to engage in an effective virtual exchange via the upper band.
Such a mechanism is expected to restore the attractive interaction between the electron and the magnon, thereby facilitating the re-formation or stabilization of the spin polaron near the critical point.
Therefore, we additionally incorporate the inter-band mixing by including the dispersive band shown in Fig.\ref{fig1}(c) in our PED calculation (see Supplemental Material) and present the results in Fig.\ref{fig3}(b).
Comparing to Fig.\ref{fig3}(a), the scatterings to the upper band dramatically facilitates spin polaron formations in that: (a) at $\delta=\delta_{c}$, the spin polaron emerges out of the $e_{\downarrow}\sigma^{-}_{a}$ continuum at $q=0$, being the global minimum of the excitation, as shown in the red dashed block; (b) the binding energies of spin polarons are universally enhanced.
These results are consistent with those reported in Ref.\cite{Prembabu2025}.
To resolve the quasiparticle embedded within the continuum or situated in the high-energy region, we calculate the spectral function with $\beta^{\dagger}_{i}=\mathcal{P}c^{\dagger}_{iA\downarrow}c_{iA\uparrow}c^{\dagger}_{iB\downarrow}\mathcal{P}$, and show the results in Fig.\ref{fig3}(c).
At $\delta=\delta_{c}$, we find that the spectral weight remains largely concentrated at the bottom edge of the $e_{\downarrow}\sigma^{-}_{a}$ continuum, signifying the persistence of the spin polaron character.
Apart from the low-energy spin polarons, we observe a dispersive band with large weight embedded within or right below (around $E \approx 0.8$) the $e_{\downarrow}\sigma^{-}_{o}$ continuum and a weak dispersive band (around $E \approx 0.9$) above the continuum at $\delta=1.0$ and $\delta_{c}$, which indicate the formation of high-energy spin polarons via the binding of a bare electron to an optical magnon.
At $\delta=1.5$, both the low-energy spin polaron around $q=\pi$ and the high-energy spin polaron below the $e_{\downarrow}\sigma^{-}_{o}$ continuum can be identified from the spectral function, though the high-energy spin polaron carries with a small spectral weight. Due to the overlap between the $e_{\downarrow}\sigma^{-}_{o}$ and $e_{\downarrow}e_{\downarrow}h_{\uparrow}$ continua, no high-energy spin polaron above the $e_{\downarrow}e_{\downarrow}h_{\uparrow}$ continuum is observed at $\delta=1.5$.
In the Supplemental Material, we present a more detailed evolution of the spin polaron by fine-tuning $\delta$, where multiple branches of spin polarons are found due to the bonding and anti-bonding between the $e_{\downarrow}$ and $\sigma^{-}_{a}/\sigma^{-}_{o}$ excitations.


\begin{figure}[htbp]
    \centering
    \includegraphics[width=\linewidth]{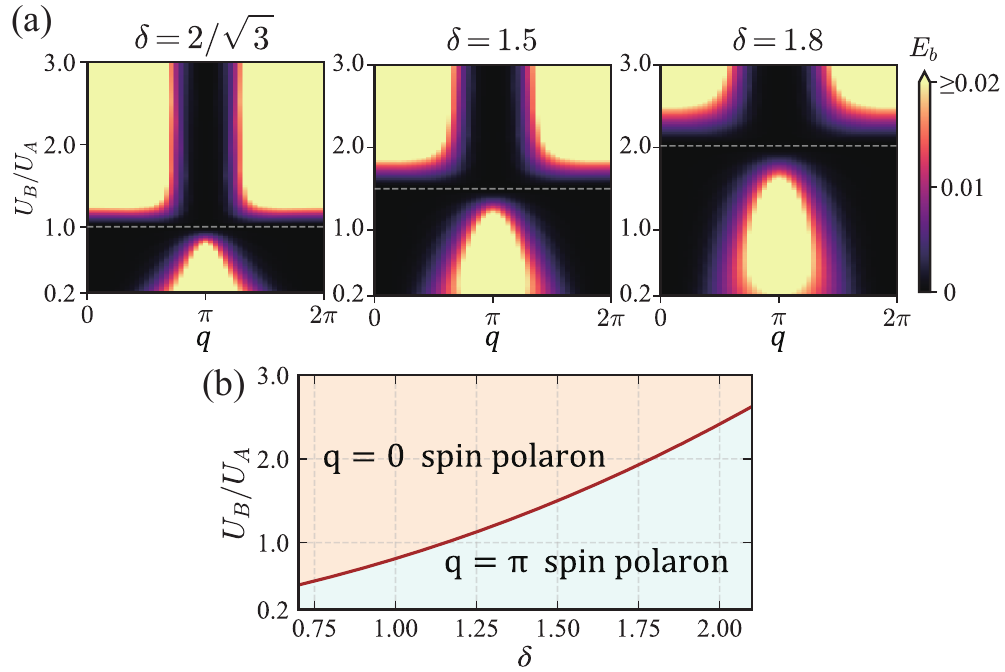}
    \caption{Relationship between the Hartree band and the spin polaron.
    (a) The binding energy of the low-energy spin polaron is presented over the entire BZ and for different $U_B/U_A$ with $\delta=2/\sqrt{3},1.5$ and $1.8$.
    The white dashed line marks the critical value of $U_B/U_A$ for a fixed  $\delta$ at which the Hartree band becomes dispersionless.
    (b) shows the parameter regions of the $q=0$ and $q=\pi$ spin polarons, which are separated by the line at which the Hartree band becomes dispersionless with the relation $U_{B}/U_{A} = \delta/(\sqrt{\delta^{2}+4}-\delta)$.
    }
    \label{fig4}
\end{figure}
Fig.\ref{fig4}(a) shows the binding energy $E_{b}(q)$, defined as the energy gap between the low-energy spin polaron and the bottom of the $e_{\downarrow}\sigma^{-}_{a}$ continuum, in the entire BZ space by varying both $\delta$ and $U_{B}/U_{A}$. The light regions denote the formation of a stable spin polaron, whereas the dark regions indicate that the spin polaron approaches to or merges into the continuum. The low-energy spin polaron first appears around $q=\pi$, and turns to be around $q=0$ with the increase of $U_{B}/U_{A}$. 
We note that the binding energy of the low-energy spin polaron vanishes around the critical parameter regime[denoted by the white dashed line in Fig.\ref{fig4}(a)], where the Hartree band becomes dispersionless, indicating the crossover of spin polarons from $q=\pi$ to $q=0$.
Then a comprehensive study of low-energy spin polaron formation over an extended range of $U_{B}/U_{A}$ and $\delta$ values are summarized in Fig. \ref{fig4}(b).
The crossover of spin polarons is around the critical value satisfying $U_{B}/U_{A} = \delta/(\sqrt{\delta^{2}+4}-\delta)$ at which the Hartree band becomes dispersionless.
This coincidence, together with the similar dispersion followed by both the low-energy spin polaron and the interaction-induced Hartree band, confirms the essential role of the Hartree dispersion in the determination of the spin polaron's momentum dependence.

\begin{figure}[htbp]
    \centering
    \includegraphics[width=\linewidth]{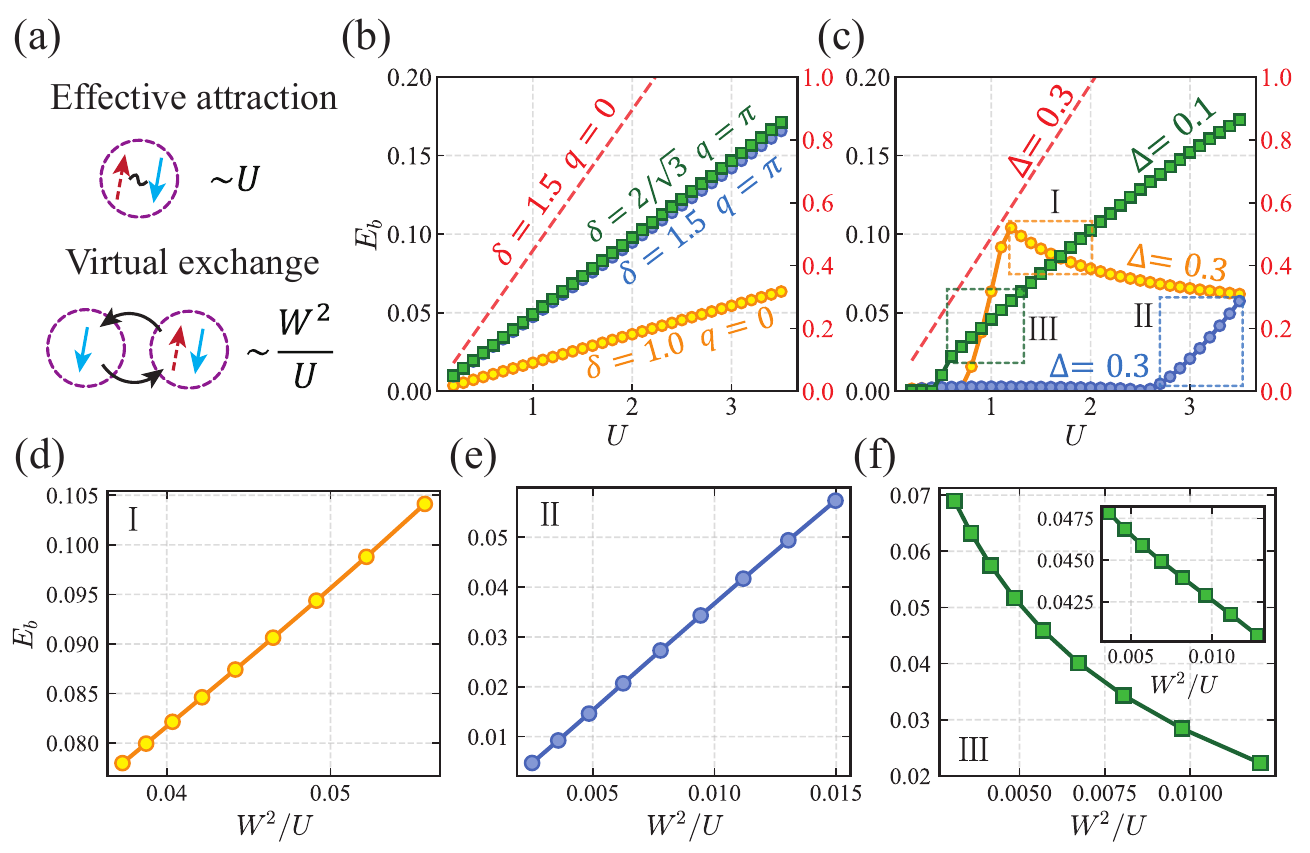}
    \caption{Binding mechanism of quasiparticles.
    (a) Sketch of the two binding mechanisms---the effective attraction and the virtual exchange.
    (b) The binding energies of acoustic magnon (red), low-energy $q=0$ (yellow), $q=\pi$ (blue) and high-energy spin polaron (green) at the corresponding $\delta$ and momentum $q$ by varying $U$.
    The red dashed line is plotted against the right y axis and the other three against the left.
    (c) The binding energies after introducing flat-band dispersion $\Delta$ (see text).
    The high-energy spin polaron is considered at $\Delta=0.1$ and the other three at $\Delta=0.3$. $\delta$'s are the same as those in (b).
    (d-f) The binding energies in the corresponding dashed blocks {\rm I, II, III} in (c) with respect to $W^2/U$, where $W$ is the width of the Hartree dispersion.
    Inset in (f) shows the binding energy of the high-energy spin polaron by varying $\Delta$ from 0.08 to 0.15 while retaining $U=1$.
    All panels have $U_{A}=U_{B}=U$.
    }
    \label{fig5}
\end{figure}

\textit{Binding mechanism.---}
Conventionally, the formation of a spin polaron is attributed to the virtual exchange between a bare electron and a magnon, leading to a binding energy proportional to $t^2/U$.
In the case of flat-band systems, another mechanism via the direct effective attraction is also expected to play a role [sketched in Fig.\ref{fig5}(a)].
To elucidate their contribution, we calculate the binding energies, which are indicated by the arrows in Fig.\ref{fig2}(a) for the acoustic magnon, and Fig.\ref{fig3}(a) for low-energy $q=0$, $q=\pi$ and high-energy spin polaron, and present the results as a function of $U$ in Fig.\ref{fig5}(b).
It shows that all binding energies exhibit a linear dependence on $U$, which seems to indicate that the formation of both magnons and spin polarons arises from the effective attraction.
This counter-intuitive scaling can be rationalized by the quenched kinetics of the flat-band system, where $U$ becomes the only energy scale after projection.
Consequently, the interaction-induced Hartree bandwidth $W$ scales with $U$ (unlike $W \propto t$ in conventional systems), so that the virtual exchange contribution ($\sim W^2/U$) also effectively follows a linear $U$-dependence.
To distinguish between these two mechanisms, we introduce an additional energy scale $\Delta$ by modifying the chemical potential in sublattice $B$ from $t\delta^2 c^{\dagger}_{iB}c_{iB}$ to $t(1+\Delta)\delta^2 c^{\dagger}_{iB}c_{iB}$.
The resulting binding energies are presented in Fig.~\ref{fig5}(c).
The binding energy of magnon retains a linear $U$-dependence [red dashed line in Fig.~\ref{fig5}(c)], indicating that its binding arises from effective attraction.
In contrast, the binding energies of the three spin polarons clearly deviate from the linear $U$-dependence.
We re-plot their binding energies as a function of $W^2/U$ in Fig.\ref{fig5}(d)-(f), focusing on the typical curve segments within the dashed blocks in Fig.\ref{fig5}(c). 
As shown in Fig.\ref{fig5}(d) and (e), the binding energies of the low-energy $q=0$ and $q=\pi$ spin polarons show a linear $W^2/U$-dependence, unambiguously confirming that they arise from the virtual exchange mechanism.
Conversely, Fig.\ref{fig5}(f) demonstrates that the binding energy of the high-energy spin polaron does not scale linearly with $W^2/U$ as $U$ is varied, indicating that its binding mechanism does not arise solely from the virtual exchange but also involves an effective attraction, {\it i.e.}, the binding energy is proportional to $\alpha U + \beta W^2/U$.
The virtual exchange contribution is demonstrated by the linear $U$-dependence in the inset of Fig.\ref{fig5}(f), which is obtained varying $\Delta$ while fixing $U$ to keep the effective attraction constant.
Overall, our work reveals a rich hierarchy of binding mechanism, emerging from the non-trivial interplay between strong correlations and residual kinetics of quasiparticle in the flat-band system.


\textit{Discussion.---}
In this work, we present a comprehensive study of spin polarons in flat-band ferromagnetism over the entire momentum and energy space, and elucidate the nature of these quasiparticles by exploring their binding mechanism.
Although our work focuses on the Mielke-Tasaki model, the conclusions apply generally to systems with a flat band and nearby dispersive bands. Therefore, we expect spin polarons to emerge in moir\'e materials—such as twisted $\text{MoTe}_2$ and $\text{WSe}_2$, where flat-band ferromagnetism has been observed~\cite{Park2023,Xu2023,Anderson2023,Cai2023,Reddy2023,Foutty2024,Pack2024,Sheng2024}.
Realizing spin polarons in these systems would provide an ideal platform for exploring potential electromagnetic engineering applications.
Our work also offers theoretical insights.
By progressively doping the itinerant flat-band ferromagnetism with spin-down electrons and spin-up holes at a 2:1 ratio, a novel phase is expected to emerge via the crystallization of spin polarons.
Such a spin polaron crystal would manifest strong spin-charge locking within the itinerant flat-band system.
Furthermore, a comprehensive understanding of the spin polaron's internal structure and dynamics may provide important microscopic insights into the unconventional superconductivity driven by spin fluctuations in moir\'e materials.
Noticing that the unconventional superconductivity has recently been observed in twisted $\text{WSe}_2$\cite{Guo2025,Xia2025,Fischer2025,Qin2025} and $\text{MoTe}_2$\cite{Xu2025a,Xu2025,Kim2025,Tuo2025}, the realization of the spin polaron mediated superconductivity\cite{Wang2025} in these materials deserves further investigation.
In short, our findings will deepen the understanding of spin polarons in flat-band systems, foster experimental exploration in moir\'e materials, and stimulate further investigations into exotic quantum phases and superconducting pairing mechanisms.

This work was supported by National Key Projects for Research and Development of China (Grant No. 2021YFA1400400), the National Natural Science Foundation of China (Grant No. 12434005, No. 12574158 and No.12550405), and the Natural Science Foundation of Jiangsu Province (Grant No. BK20233001).

\bibliography{ref}

\clearpage
\onecolumngrid

\begin{center}
  \textbf{Supplemental Material for ``Interaction-driven spin polaron in itinerant flat-band ferromagnetism''}
\end{center}

\setcounter{equation}{0}
\setcounter{figure}{0}
\setcounter{table}{0}
\setcounter{page}{1}
\setcounter{section}{0}

\renewcommand{\theequation}{S\arabic{equation}}
\renewcommand{\thefigure}{S\arabic{figure}}
\renewcommand{\thetable}{S\arabic{table}}
\renewcommand{\thesection}{S-\Roman{section}}

\section{Mielke-Tasaki model}
The Mielke-Tasaki model exhibits an exact flat band by fine tuning the hopping parameters to satisfy the destructive interference criterion.
Here we does not begin with the simple diagonalization of the single-particle Hamiltonian, but more intuitively show how the compact localized state can appear by the destructive interference.
\begin{figure}[h]
    \centering
    \includegraphics[width=0.5\linewidth]{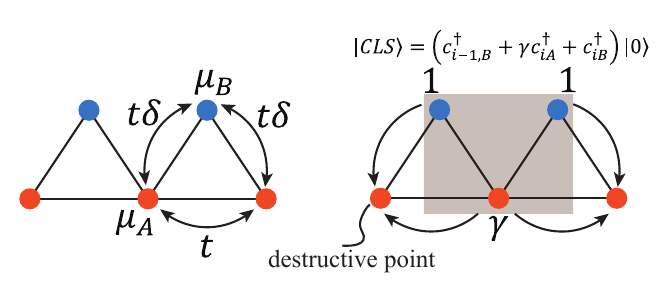}
    \caption{
    \textbf{Sketch of the compact localized state.}
    }
    \label{sm_fig_cls}
\end{figure}

As shown above, the shaded region shows the compact localized state which has:
\begin{equation}
    \ket{\text{CLS}} = ( c^{\dagger}_{i-1,B} + \gamma c^{\dagger}_{iA} + c^{\dagger}_{iB} ) \ket{0}
    \label{eqs1}
\end{equation}

In order for this state to be localized, it should not move out of this three sites, so that $t\delta \times 1 + t \times \gamma = 0$, which has $\gamma = -\delta$.
And this state should be the eigen state of the Hamiltonian, so that $H\ket{\text{CLS}} = E\ket{\text{CLS}}$, which leads to:

\begin{equation}
    E = \frac{\mu_{B} + (-\delta) \times t\delta}{1} = \frac{(-\delta) \times \mu_{A} + 2 \times t\delta}{-\delta}
    \label{eqs2}
\end{equation}

one can choose $\mu_{A} = 2t, \mu_{B} = t\delta^{2}$, and the eigen energy is $E=0$.
After properly choose $\mu_{A/B}$, this compact localized state becomes the eigen state of the Hamiltonian and the set consisting of these states with the label $i$ ranging from 1 to N span the space of the flat band.
This intuitive picture elucidates that the essence of the flat band is due to the fine tuning of the hopping parameters to host such compact localized state which can not hop to other sites, resembling an atomic chain with nearest hopping $t=0$.

\clearpage
\section{Projected exact diagonalization and inter-band mixing}
For an electron system with the kinetics $H_{0} = \sum_{k\mu \sigma} \varepsilon^{\sigma}_{k\mu} \alpha^{\dagger}_{k\mu \sigma}\alpha_{k\mu \sigma}$ ($\mu$ the band index) and the Hubbard term $H_{U} = \sum_{ni} U_{n}n^{\downarrow}_{ni}n^{\uparrow}_{ni}$ ($n=A,B$), if the band gap is
much larger than the Hubbard $U$ while the lower band is almost flat, the ground state would exhibit an itinerant ferromagnetism at $1/4$ filling to minimize the interaction energy, $\ket{\text{GS}}=\prod_{k\in 1BZ} \alpha^{\dagger}_{kl\uparrow}\ket{0}$.

In spirit of the single mode approximation, we can include the scattering to the upper band by enlarging the projection space to from $\mathcal{P}=\sum_{\{k_{1},k_{2},k_{3}\}} \ket{k_{3},k_{2};k_{1}}\bra{k_{3},k_{2};k_{1}}$ to $\mathcal{P}=\sum_{\{k_{1},k_{2}\mu_{2},k_{3}\mu_{3}\}} \ket{k_{3}\mu_{3},k_{2}\mu_{2};k_{1}}\bra{k_{3}\mu_{3},k_{2}\mu_{2};k_{1}}$, where $\mu=l,u$ sums over the lower flat band and the upper dispersive band, $\ket{k_{3}\mu_{3},k_{2}\mu_{2};k_{1}} = \alpha^{\dagger}_{k_{3}\mu_{3}\downarrow} \alpha^{\dagger}_{k_{2}\mu_{2}\downarrow} \alpha_{k_{1}l\uparrow} \ket{\text{GS}}$.

the $e_{\downarrow}e_{\downarrow}h_{\uparrow}$ excitation can be obtained by the following equations:

\begin{equation}
    \mathcal{P}H_{0}\mathcal{P}\ket{k_{3}\mu_{3},k_{2}\mu_{2};k_{1}} = \{ \varepsilon^{\downarrow}_{\mu_{3}}(k_{3}) + \varepsilon^{\downarrow}_{\mu_{2}}(k_{2}) -
    \varepsilon^{\uparrow}_{l}(k_{1}) \} \ket{k_{3}\mu_{3},k_{2}\mu_{2};k_{1}}
    \label{eqS3}
\end{equation}

\begin{equation}
\begin{split}
    \mathcal{P}H_{U}\mathcal{P} &\ket{k_{3}\mu_{3},k_{2}\mu_{2};k_{1}} = \frac{1}{N} \sum_{nkk'} M^{\uparrow,ll}_{n}(k',0)\{ M^{\downarrow,\mu \mu_{2}}(k_{2},0) \ket{k_{3}\mu_{3},k_{2}\mu;k_{1}} + M^{\downarrow,\mu \mu_{3}}_{n}(k_{3},0)\ket{k_{3}\mu,k_{2}\mu_{2};k_{1}} \} \\ &- \frac{1}{N}\sum_{nkq} M^{\uparrow,ll}_{n}(k_{1}+q,-q) \{ M^{\downarrow,\mu \mu_{2}}_{n}(k_{2},q) \ket{k_{3}\mu_{3},(k_{2}+q)\mu;k_{1}+q} + M^{\downarrow,\mu \mu_{3}}_{n}(k_{3},q) \ket{(k_{3}+q)\mu,k_{2}\mu_{2};k_{1}+q} \}
    \label{eqS4}
\end{split}
\end{equation}

where $M^{\sigma,\mu\nu}_{n}(k,q) = U^{\sigma*}_{n\mu}(k+q)U^{\sigma}_{n\nu}$, $U^{\sigma}_{n\mu}(k)$ is the unitary transformation between $\alpha_{k\mu\sigma}$ and $c_{kn\sigma}$. When the band width of the lower band vanishes, Eq.\ref{eqS3} is zero; while introducing the extra chemical potential $t\delta c^{\dagger}_{iB}c_{iB}$ in the main text, Eq.\ref{eqS3} should be included.

\clearpage
\begin{figure}
    \centering
    \includegraphics[width=\linewidth]{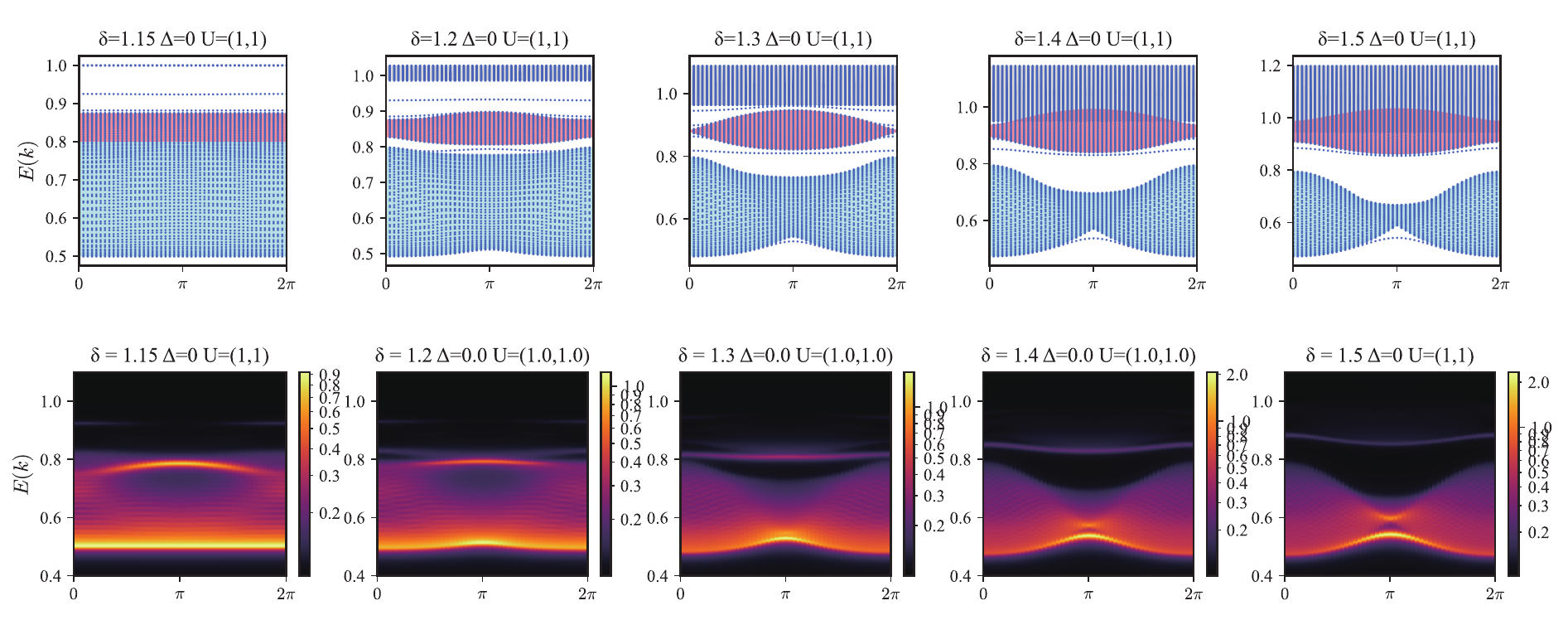}
    \caption{
    \textbf{Variation of spectrum and spectral density of spin polaron.}
    By varying $\delta$ from $2/\sqrt{3}$ $(\approx 1.15)$ to 1.5, multiple branches of spin polaron is observed both by energy and spectral density.
    At $\delta=\text{1.3}$, four branches locate symmetrically around the $e_{\downarrow}\sigma^{-}_{o}$ continuum, indicating the bonding and anti-bonding of the bare electron and the optical magnon.
    By the evolution of the spectral density, one can identify that the bright branch around $E=0.8$ at $\delta=2/\sqrt{3}$ is the spin polaron emerged from the binding of the bare electron to the optical magnon.
    }
    \label{sm_fig1}
\end{figure}

\clearpage
\begin{figure}
    \centering
    \includegraphics[width=\linewidth]{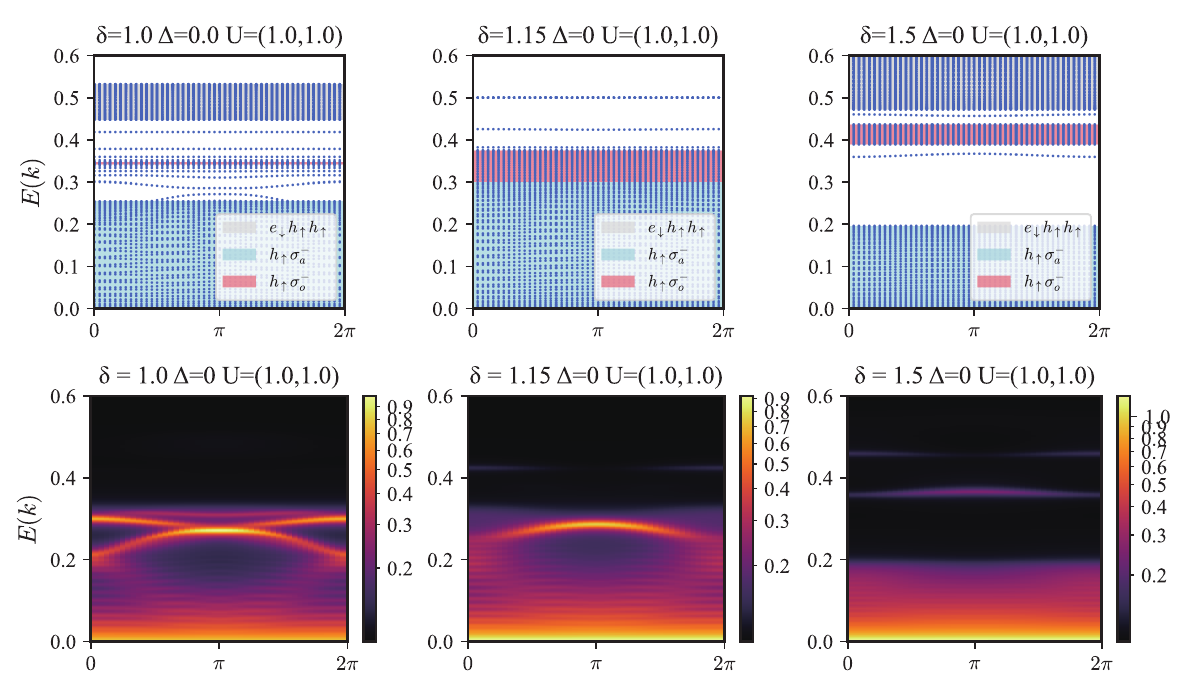}
    \caption{
    \textbf{Spectrum and spectral density of the $e_{\downarrow}h_{\uparrow}h_{\uparrow}$ excitation.}
    The low-energy spin polaron always merge into the $e_{\downarrow}\sigma^{-}_{a}$ continuum, which shows that the non-vanishing Hartree dispersion plays an important role for the emergence of the low-energy spin polaron.
    Another interesting thing is that at $\delta=1$ where the optical magnon dispersion is almost flat [see Fig.2(a) in the main text], there are multiple branches of dispersion locating around the $e_{\downarrow}\sigma^{-}_{o}$ continuum.
    This indicates novel binding patterns between the bare hole and the optical magnon, which are not characterized by the quasiparticle ``probe'' $\beta^{\dagger}_{i}=\mathcal{P}c^{\dagger}_{iA\downarrow}c_{iA\uparrow}c^{\dagger}_{iB\downarrow}\mathcal{P}$, as the dispersions above the $h_{\uparrow}\sigma^{-}_{o}$ continuum show no spectral density.
    }
    \label{sm_fig2}
\end{figure}

\end{document}